\documentclass[aps,prl,showpacs,floatfix,superscriptaddress,twocolumn]{revtex4-1}
\usepackage{graphicx}
\usepackage{amsmath,amssymb,amsthm}
\usepackage{times}
\usepackage{bm}
\usepackage[dvips]{color}

\newcommand{\be}{\begin{eqnarray}}
\newcommand{\ee}{\end{eqnarray}}

\def\beq{\begin{equation}}
\def\eeq{\end{equation}}

\def\kini{|\psi_I\rangle}
\def\bini{\langle\psi_I|}
\def\a{\alpha}
\def\Abar{\overline{A}}
\def\Aaa{A_{\a\a}}
\def\Aab{A_{\a\b}}
\def\b{\beta}
\def\Ca{C_\a}
\def\cg{{\rm cg}}
\def\Ebar{\bar E}

\def\H{{\cal H}}
\def\la{\langle}
\def\mc{{\rm mc}}
\def\ra{\rangle}

\def\ltwid{\mathrel{\raise.3ex\hbox{$<$\kern-.75em\lower1ex\hbox{$\sim$}}}}
\def\gtwid{\mathrel{\raise.3ex\hbox{$>$\kern-.75em\lower1ex\hbox{$\sim$}}}}

\newcommand{\expv}[1]{\langle #1 \rangle}	

\begin{document}

\title{Alternatives to Eigenstate Thermalization}

\author{Marcos Rigol}
\affiliation{Department of Physics, Georgetown University, Washington, D.C. 20057, USA}

\author{Mark Srednicki}
\affiliation{Department of Physics, University of California, Santa Barbara, California 93106, USA}

\begin{abstract}
An isolated quantum many-body system in an initial pure state will come to thermal equilibrium 
if it satisfies the {\em eigenstate thermalization hypothesis} (ETH). We consider alternatives to 
ETH that have been proposed. We first show that von Neumann's quantum ergodic theorem 
relies on an assumption that is essentially equivalent to ETH. We also investigate whether, 
following a sudden quench, special classes of pure states can lead to thermal behavior in systems 
that do not obey ETH, namely, integrable systems. We find examples of this, but only for initial 
states that obeyed ETH before the quench.
\end{abstract}

\maketitle

{\it Introduction and summary of ETH:} Consider an isolated quantum $N$-body system with Hamiltonian 
$\hat{H}$.  Let $|\a\ra$ denote an eigenstate of $\hat{H}$ with eigenvalue $E_\a$, and let $\hat{A}$ denote 
a few-body observable. The {\em eigenstate thermalization hypothesis} (ETH) states that 
(1) the diagonal matrix elements $\Aaa=\la\a|\hat{A}|\a\ra$ change slowly with the state, with the difference between
neighboring values $A_{\alpha+1,\alpha+1}-\Aaa$ exponentially small in $N$, and (2)
that the off-diagonal matrix elements $\Aab=\la\a|\hat{A}|\b\ra$, $\a\ne\b$, are themselves
exponentially small in $N$ \cite{deutsch_91,srednicki_94}.  
ETH is suggested by various results in quantum chaos theory (in particular, Shnirelman's theorem 
\cite{shnirelman_74} and Berry's random-wave conjecture \cite{berry_77}) for systems that have a chaotic 
classical limit. ETH has been verified numerically in a wide variety of quantum many-body systems that are 
sufficiently far (in parameter space) from points of integrability 
\cite{rigol_dunjko_08,rigol_09a,santos_rigol_10a,neuenhahn_marquardt_10},
but it certainly does not hold in systems that are integrable or near integrable 
\cite{rigol_dunjko_08,rigol_09a,santos_rigol_10a,neuenhahn_marquardt_10,cassidy_clark_11}.

Let $|\psi(\tau)\ra$ be the quantum state of the system at time $\tau$, given by the evolution of some 
initial state $\kini=\sum_\a \Ca|\a\ra$,
\beq
|\psi(\tau)\ra = e^{-i\hat{H} \tau/\hbar}\kini=\sum_\a \Ca e^{-iE_\a \tau/\hbar}|\a\ra,
\label{psi}
\eeq
with $\sum_\a |\Ca|^2=1$.  The energy of the system is $\Ebar = \sum_\a |\Ca|^2 E_\a$, and the quantum 
energy uncertainty is $\Delta E$, where $(\Delta E)^2=\sum_\a |\Ca|^2 (E_\a-\Ebar)^2$.  We assume that 
$\Delta E$ is algebraically small in $N$,
(e.g., $\Delta E\sim N^{-1/2}\Ebar$), as is the case for any state of a macroscopic system that could be 
realistically prepared in a laboratory. (This is also true for sudden quenches between Hamiltonians with 
short-range interactions \cite{rigol_dunjko_08}.) The time-dependent expectation value of $\hat{A}$ is
\be
\la \hat{A}(\tau)\ra &=& \la\psi(\tau)|\hat{A}|\psi(\tau)\ra 
\label{A(tau)} \\
&=&\sum_\a |\Ca|^2\Aaa + \sum_{\a\ne\b}C^*_\a C_\b e^{i(E_\a-E_\b)\tau/\hbar}\Aab,
\nonumber 
\ee
and the long-time average of $\la \hat{A}(\tau)\ra$ is 
\beq
\Abar = \lim_{\tau\to\infty} \frac1\tau\int_0^\tau dt\,\la \hat{A}(\tau)\ra = \sum_\a |\Ca|^2\Aaa,
\label{Abar}
\eeq
in the absence of degeneracies (which are not expected to occur in chaotic systems without extra 
symmetries \cite{bohigas_giannoni_84}). 
The right-hand side of Eq.~\eqref{Abar} effectively sums $|C_\a|^2\Aaa$ over an energy window of width $\Delta E$ 
that is centered on $\Ebar$. According to ETH, $\Aaa$ is approximately constant over this window.  Thus,  
up to algebraically small corrections, the right-hand side of Eq.~\eqref{Abar} has the same value as the 
microcanonical average of $\hat A$ over the same window, and is independent of the detailed pattern of values 
taken by the $|C_\a|^2$ coefficients. Thus ETH results in the equality of time averages and thermal averages for 
a very broad class of initial states.

From ETH, each term in the second sum in Eq.~\eqref{A(tau)} is exponentially small in $N$. However, the number 
of terms in this sum is exponentially large, and so, if the phases line up coherently, the second sum 
can rival the first.  If this happens at one particular time (say, $\tau=0$), it will fail at sufficiently 
large later times, as the time-dependent phases go out of alignment.  This dephasing mechanism accounts 
for the approach to thermal equilibrium for an initially out-of-equilibrium state 
\cite{srednicki_94,rigol_dunjko_08}.

{\it The quantum ergodic theorem:} In 1929, von Neumann proved a mathematical result which has been 
dubbed the quantum ergodic theorem (QET) \cite{vonneumann_10}. An exegesis of it has been given by 
Goldstein {\it et al.}~(hereafter GLTZ) \cite{goldstein_lebowitz_10}. GLTZ summarize QET, 
or ``normal typicallity'' as it has been more recently known, as follows: ``for a typical finite family of 
commuting macroscopic observables, every initial wave function from a microcanonical energy shell so 
evolves that for most times $\tau$ in the long run, the joint probability distribution of these observables 
obtained from $|\psi(\tau)\ra$ is close to their microcanonical distribution'' \cite{goldstein_lebowitz_10}. 
More specifically, QET states that 
$\la \hat{A}(\tau)\ra$ will be close to the microcanonical average of $\hat{A}$ in the following sense: 
$|\la \hat{A}(\tau)\ra-\expv{\hat{A}}_\mc|^2 < \epsilon^2 \expv{\hat{A}^2}_\mc$ for all but a fraction 
$\delta$ of times $t$, where the subscript mc denotes the microcanonical average over the energy window 
of all states with nonzero $\Ca$, and $\epsilon$ and $\delta$ are small numbers.  
The proof of the theorem requires that all energy eigenvalue differences $E_\a-E_\b$ be nondegenerate,
and an additional condition that is deemed technical by GLTZ, their Eq.~(17). Here we point out that 
this condition is equivalent to ETH. Hence, von Neumann's proof of QET relies on ETH \cite{note1}.

We follow the exposition of GLTZ, and consider the system to have a Hilbert space $\H$ with
an exponentially large but finite dimension $D$.  We focus on a single observable $\hat{A}$.  GLTZ 
partition the Hilbert space into ``macrospaces'' $\H_\nu$, with dimension $d_\nu$.  Each $\H_\nu$ is 
spanned by the eigenstates of $\hat{A}$ with eigenvalue $a$ in a particular range centered on a value $a_\nu$.
Let $\hat{P}_\nu$ be the projection operator onto $\H_\nu$, and consider its
energy-basis matrix elements $\la\a|\hat{P}_\nu|\b\ra$.  The condition
that must be assumed to prove QET is that, for each $\nu$, the off-diagonal elements ($\a\ne\b$)
must be exponentially small, and the diagonal elements $\la\a|\hat{P}_\nu|\a\ra$
must be exponentially close to $f_\nu=d_\nu/D$, the fraction of states in $\H_\nu$
\cite{vonneumann_10,goldstein_lebowitz_10}.

We now argue that this condition is effectively equivalent to ETH. Consider the operator 
$\hat{A}_\cg\equiv\sum_\nu a_\nu \hat{P}_\nu$.  This is a coarse-grained version of $\hat{A}$
itself, which can be written as $\hat{A}=\sum_a a|a\ra\la a|$, $\hat{A}_\cg$ is the same, but with 
the individual values of $a$ replaced by their average values in each macrospace.  
The intention of von Neumann and GLTZ is that $\hat{A}_\cg$ should be indistinguishable from
$\hat{A}$ in practical experiments.  Next, consider the energy-basis expectation value 
$\la\a|\hat{A}_\cg|\a\ra$. Since $\la\a|\hat{P}_\nu|\a\ra$ is (by the GLTZ technical condition) 
exponentially close to $f_\nu$, then from the definition of $\hat{A}_\cg$ we get
$\la\a|\hat{A}_\cg|\a\ra = \sum_\nu a_\nu f_\nu$, up to exponentially small corrections.
Up to the additional small errors introduced by the coarse graining (which are assumed to be
negligible), this last expression is equal to the trace of $\hat{A}$, which in this simplified 
model is to be identified with the microcanonical average of $\hat{A}$. This is equivalent to 
the ETH statement that an energy-basis expectation value is equivalent to a microcanonical 
average. Similarly, if $\la\a|\hat{P}_\nu|\b\ra$ is exponentially small for $\a\ne\b$, so is 
$\la\a|\hat{A}|\b\ra$, which is the other key part of ETH \cite{srednicki_94,rigol_dunjko_08}.  

Another way to phrase the equivalence is to note that both the GLTZ technical condition and ETH
rely on exponential smallness of the overlap between an energy eigenstate and an
eigenstate of an observable $\hat{A}$ that exhibits thermal behavior.
As already noted, ETH can be justified by various results from quantum chaos theory.
Thus, ETH provides a physical basis for the technical condition needed by QET. 
This results in a unification of two formerly disparate schools of thought on the foundations 
of statistical mechanics.  

{\it Quantum quenches:} As mentioned in the introduction, ETH has been shown to be satisfied
in a variety of nonintegrable quantum systems. It has 
been found to break down only as one approaches integrable points 
\cite{rigol_09a,santos_rigol_10a,neuenhahn_marquardt_10}, or in special 
regimes that are dominated by finite size effects, {\it e.g.}, close to the atomic limit 
\cite{roux_10,rigol_santos_10}. Thermalization itself has been shown to be robust
in nonintegrable systems after a (sudden) quench, once again, failing to occur close to integrable 
points \cite{manmana_wessel_07,rigol_09a,banuls_cirac_11} or the atomic limit 
\cite{kollath_lauchli_07,rigol_santos_10}, and in localized disordered systems 
\cite{gogolin_muller_11}. Here, by (sudden) quench, we mean that the system is prepared in an eigenstate
of some initial Hamiltonian (not necessarily the ground state) and then at $\tau=0$ the 
Hamiltonian is changed. 

We now consider systems that do not satisfy ETH; in particular, we consider systems for which
$\Aaa$ varies significantly [that is, by an amount that is $O(N^0)$] with $\a$.  
It is easy to construct such systems; for example, any set of noninteracting degrees of freedom,
with $\hat{A}$ corresponding to any one- or few-body observable, is in this class.  Interacting systems 
that are integrable (with as many conserved charges as degrees of freedom) are in this class
\cite{sutherland_book_04}. Equation \eqref{Abar} still applies to such systems; but now whether 
or not $\Abar$ is close to the thermal average of $\hat{A}$ depends strongly on the initial state, 
which is specified by the $\Ca$ coefficients.  If the values of the coefficients $|\Ca|^2$ in 
Eq.~\eqref{Abar} provide an {\em unbiased sampling} of the matrix elements $\Aaa$, then we can expect
$\Abar$ to be 
algebraically 
close to the microcanonical average of $\hat{A}$ over the energy window
specified by the $\Delta E$ of the initial state. 

The above scenario, however, does not occur
in quenches between integrable systems, i.e., when the initial state is an eigenstate of 
an integrable system, and the time evolution is studied after changing some parameters in the 
Hamiltonian while keeping the system integrable. Studies of several models have shown that
$\Abar$ remains different from the thermal expectation as one approaches the thermodynamic 
limit \cite{rigol_dunjko_07,cazalilla_06,calabrese_cardy_07a,
cramer_dawson_08,barthel_schollwock_08,eckstein_kollar_08,kollar_eckstein_08,rossini_silva_09,
iucci_cazalilla_09,fioretto_mussardo_10,mossel_caux_10,cassidy_clark_11,calabrese_essler_11,
cazalilla_iucci_11,rigol_fitzpatrick_11}. Even some special initial states that were seen to 
lead to $\Abar$ similar to the ones predicted in thermal equilibrium 
\cite{rigol_dunjko_07,rossini_silva_09,cassidy_clark_11}, have been recently shown 
not to result in the thermalization $\hat{A}$ in the thermodynamic limit 
\cite{calabrese_essler_11,rigol_fitzpatrick_11}.

Here we identify a class of initial states that leads to thermal behavior after a quench to an integrable point.  
The initial states we consider are eigenstates of an initial Hamiltonian $\hat{H}_I$ that is nonintegrable.
This Hamiltonian is constructed by breaking the integrability of the final Hamiltonian $\hat{H}_F$; i.e., 
we set $\hat{H}_I=\hat{H}_F+{}$``integrability breaking terms.'' The idea here is that the 
integrability breaking terms in the initial Hamiltonian generate eigenstates that are 
unbiased combinations of eigenstates of the integrable (final) Hamiltonian. This is what 
leads to chaotic behavior as one departs from an integrable point \cite{santos_rigol_10a}, 
and ultimately allows ETH to be valid in nonintegrable systems.  Hence, such initial states enable 
the desired unbiased sampling that does not occur in quenches between integrable systems.

In order to show that this is indeed the case, we have studied one-dimensional lattice 
systems of hard-core bosons and spinless fermions with the Hamiltonian
{\setlength\arraycolsep{0.5pt}
\begin{eqnarray}
&& H =\sum_{j=1}^{L} \Bigl[ -t \Bigl(\hat{c}_j^{\dagger} \hat{c}^{}_{j+1} + \textrm{H.c.} \Bigr)
 -t' \Bigl( \hat{c}_j^{\dagger} \hat{c}^{}_{j+2} + \textrm{H.c.}\Bigr)+
\nonumber \\
&& 
V \Bigl(\hat{n}_j -{\textstyle\frac{1}{2}} \Bigr)\Bigl(\hat{n}_{j+1} -{\textstyle\frac{1}{2}}\Bigr)
+V' \Bigl(\hat{n}_{j} -{\textstyle\frac{1}{2}}\Bigr)\Bigl(\hat{n}_{j+2} -{\textstyle\frac{1}{2}}\Bigr)
\Bigr],\quad
\label{bosonHam}
\end{eqnarray}
}where $\hat{c}_j^{\dagger}$ ($\hat{c}_j^{}$) stands for the creation (annihilation) operator for 
hard-core bosons and fermions, $\hat{n}_j=\hat{c}_j^{\dagger}\hat{c}_j^{}$ is the site occupation 
operator, $L$ is the number of lattice sites, and $t$ ($t'$) and $V$ ($V'$) are the nearest (next-nearest) 
neighbor hopping and interaction, respectively. The $t'$, $V'$ terms are the ones that make this Hamiltonian 
nonintegrable. We consider periodic boundary conditions, and the full diagonalization of the Hamiltonian is done 
using its translational symmetry. The filling is always taken to be $N=L/3$.

Our quench protocol is then as follows, we generate an initial state $\kini$ that is an eigenstate of 
Eq.~\eqref{bosonHam} with $t=V=1$ (this sets our energy scale), $t'=V'\neq0$, and which lies within
the sector of zero total momentum. The final Hamiltonian, after the quench, still has $t=V=1$, but we set 
$t'=V'=0$. This Hamiltonian is integrable. We then study the relaxation dynamics of various observables, 
as well as their description after relaxation, for many different initial states. 
These initial states 
are eigenstates of Hamiltonians with different values of $t',V'$, and are selected so that the system 
can have different final effective temperatures. The effective temperature $T$ is calculated using the 
standard procedure for the canonical ensemble, i.e., such that 
$E = Z^{-1} \textrm{Tr}(\hat{H}_Fe^{-\hat{H}_F/k_BT})$, 
where $E=\bini\hat{H}_F\kini$ is the energy of the time evolving state, $Z=\textrm{Tr}(e^{-\hat{H}_F/k_BT})$ 
is the partition function, and $k_B$ (set to one in what follows) is the Boltzman constant.

We have studied four observables, the kinetic energy $\hat{K}$, the interaction energy $\hat{U}$, 
the momentum distribution function $\hat{n}_k$ (which is the Fourier transform of the one-particle correlations 
$\hat{\rho}_{ij}=\hat{c}_i^{\dagger}\hat{c}_j^{}$), and the structure factor $\hat{N}_k$ 
(which the Fourier transform of the density-density correlations 
$\hat{N}_{ij}=\hat{n}_i\hat{n}_j$). $\hat{K}$ and $\hat{U}$ are local observables while $\hat{n}_k$ and 
$\hat{N}_k$ are nonlocal, and $\hat{K}$ and $\hat{n}_k$ are single-body observables while $\hat{U}$ and 
$\hat{N}_k$ are two-body observables. In all cases studied, we found a similar qualitative behavior 
in those four quantities. Hence, we will only report results for $\hat{n}_k$, which is the one with 
the closest connection to ultracold gases experiments 
\cite{kinoshita_wenger_06,hofferberth_lesanovsky_07,hung_zhang_10,trotzky_chen_11}.

\begin{figure}[!t]
\begin{center}
\includegraphics[width=0.48\textwidth,angle=0]{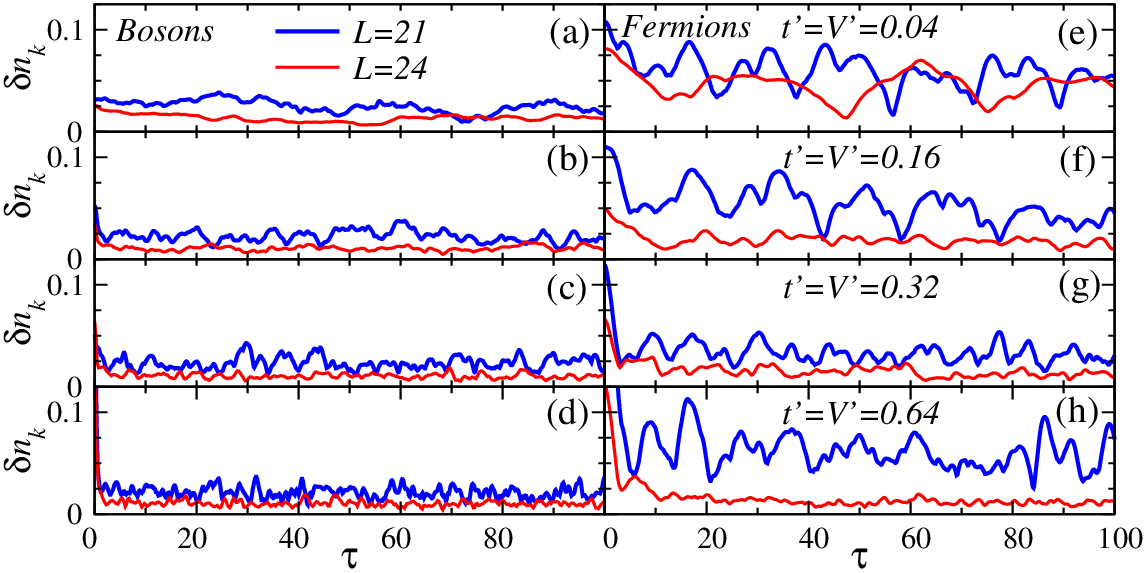}
\end{center}
\vspace{-0.6cm}
\caption{\label{nk_TimeEvol_T3.0} Time evolution of $\delta n_k$, after a sudden quench, 
for hard-core bosons (left panels) and spinless fermions (right panels) 
in lattices with $L=21$ and $L=24$ ($N=L/3$), and $T=3$. Results are presented for quenches 
from $t'=V'=0.04$ (a),(e), $t'=V'=0.16$ (b),(f), $t'=V'=0.32$ (c),(g), and
$t'=V'=0.64$ (d),(h), to $t'=V'=0$. In all cases $t=V=1$ before and after the quench.}
\end{figure}

We are first interested in understanding how, after the quench, 
observables relax (if they do) to the long-time average in Eq.~\eqref{Abar}. For $\hat{n}_k$, 
this can be conveniently quantified by calculating the normalized integrated difference
\begin{equation}
 \delta n_k(\tau)=\dfrac{\sum_k|\expv{\hat{n}_k(\tau)}-\overline{n_k}|}{\sum_k \overline{n_k}}.
\label{error}
\end{equation}

Results for $\delta n_k(\tau)$ are shown in Fig.~\ref{nk_TimeEvol_T3.0} for bosons (left column)
and fermions (right column) and for two different system sizes. For very small quenches 
[Fig.~\ref{nk_TimeEvol_T3.0}(a),(e)], the initial state has a nonzero overlap only with very 
few close-by eigenstates of the final Hamiltonian. Because of this, the long-time average is 
close to the initial $n_k$, and large oscillations (relative to the time average) can be seen 
for both bosons and fermions. By increasing the amplitude of the quench, the long-time average 
becomes increasingly different from the initial momentum distribution, and the 
time fluctuations decrease. This is because more eigenstates of the 
final Hamiltonian are involved in the dynamics and dephasing becomes more efficient. 
It is interesting to note the strikingly large finite size effects seen for the spinless 
fermions [Fig.~\ref{nk_TimeEvol_T3.0}(e)--(g)], where the dynamics changes (improves) dramatically 
by increasing the system size from $L=21$ to $L=24$ \cite{rigol_09a,santos_rigol_10a}. Overall, 
one can conclude from those results that, as the system size increases, $n_k$ relaxes to the 
long-time average rather quickly ($\tau\sim \hbar/t$), and the time fluctuations around that average 
become very small. Similar results were found for other observables and effective temperatures.

\begin{figure}[!t]
\begin{center}
\includegraphics[width=0.35\textwidth,angle=0]{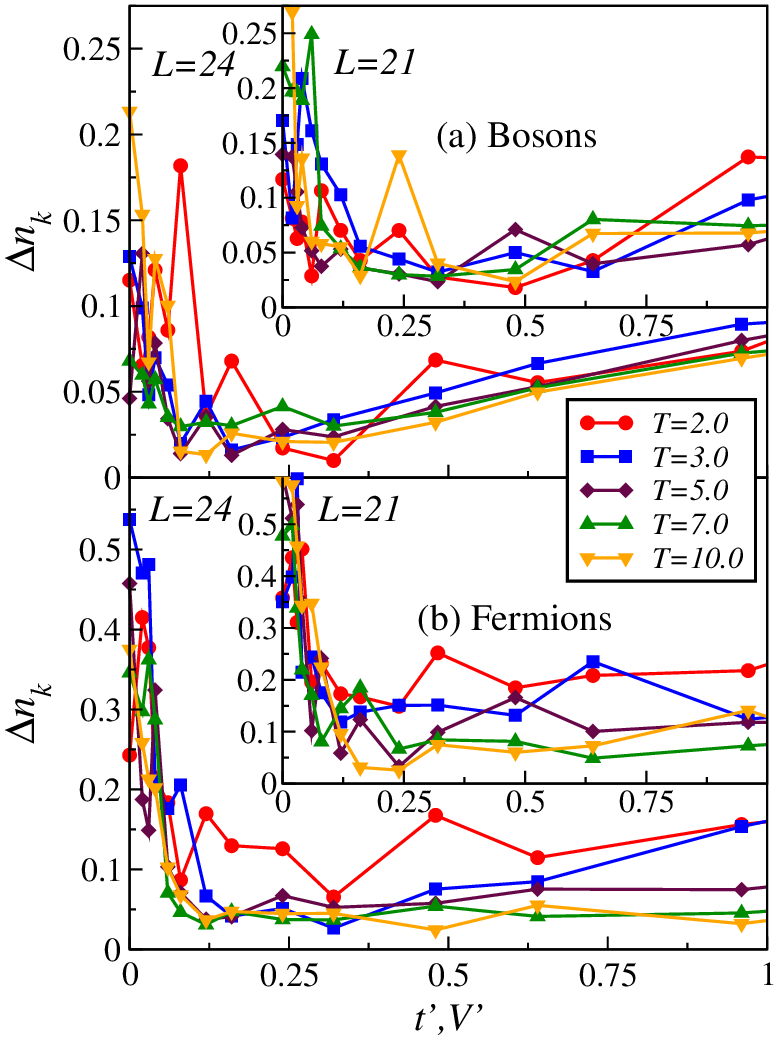}
\end{center}
\vspace{-0.6cm}
\caption{\label{nk_DiagvsMic} Relative difference between $n_k$ as predicted by the 
microcanonical ensemble and the long-time average, as a function of $t',V'$ in the 
initial Hamiltonian. Results are reported for hard-core bosons (a) and spinless fermions (b), 
for lattices with 24 sites (main panels) and 21 sites (insets). In all 
systems, we considered five different initial states that resulted in 
five effective temperatures after the quench ($T=2,\,3,\,5,\,7$, and 10 in the plots).}
\end{figure}

We then are left to check how accurate statistical ensembles are when predicting $\overline{n_k}$. 
For our small finite systems, we use the microcanonical ensemble, 
and calculate the following normalized integrated difference to quantify its accuracy
\begin{equation}
 \Delta n_k=\dfrac{\sum_k|\expv{\hat{n}_k}_\textrm{mc}-\overline{n_k}|}{\sum_k \overline{n_k}}.
\label{Error}
\end{equation}
The width $\Delta E$ of the energy window in the microcanonical ensemble is taken such that the 
results are robust to small changes of $\Delta E$ (in our systems $\Delta E\sim$0.1--0.3). 

Figure~\ref{nk_DiagvsMic} depicts our results for $\Delta n_k$, calculated for bosons (a)
and fermions (b), and for $L=24$ (main panels) and $L=21$ (insets). In that figure, it is apparent 
that $n_k$ is nonzero, and large, for small values of $t',V'$. This is true even in the absence of 
a quench, when $t'=V'=0$ in the initial and final Hamiltonian, and reflects the failure of 
ETH due to integrability. However, as the value of $t',V'$ is increased, $n_k$ 
is seen to decrease in all cases. This occurs for bosons and fermions at all 
effective temperatures and system sizes. For all observables 
and effective temperatures that we have studied, we have found that $\Delta n_k$ decreases
with increasing system size. This, together with the understanding of the role of the integrability 
breaking terms in the initial Hamiltonian, supports our claim that initial states that
satisfy ETH (eigenstates of a nonintegrable Hamiltonian) will lead to thermalization in integrable 
systems, despite the fact that the latter do not satisfy ETH.

\begin{figure}[!t]
\begin{center}
\includegraphics[width=0.48\textwidth,angle=0]{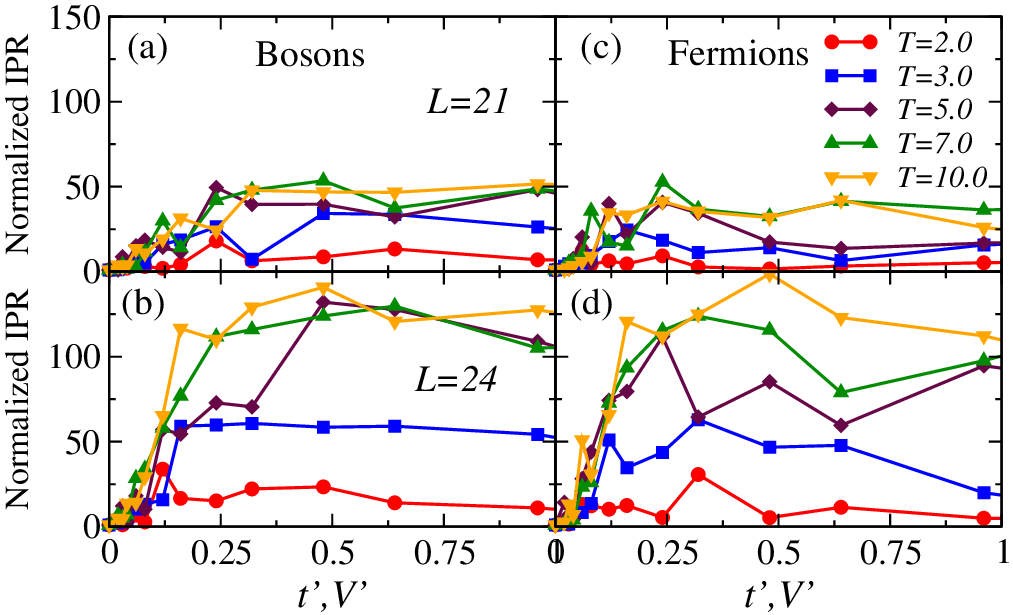}
\end{center}
\vspace{-0.6cm}
\caption{\label{NormalizedIPRvstpVp} Normalized IPR (see text) for the
same quenches depicted in Fig.~\ref{nk_DiagvsMic}. Results for hard-core bosons
(fermions) are shown in the left (right) panels, and for $L=21$ ($L=24$)
in the top (bottom) panels.}
\end{figure}

In Fig.~\ref{nk_DiagvsMic}, one can also see that, in many instances, $\Delta n_k$ for a fixed 
system size reaches a minimum value and then increases as $t'$ and $V'$ are increased. This occurs 
because, for large values of $t',V'$, the initial state starts having large overlaps with 
eigenstates outside the microcanonical energy window. Hence, even though the initial state 
is sampling more eigenstates of the final Hamiltonian, the energy density becomes broad 
and the microcanonical ensemble once again becomes a bad approximation for the long-time average.
The fact that more eigenstates of the final Hamiltonian are part of the initial state can 
be quantified by means of the inverse participation ratio (IPR)
$\textrm{IPR}=1/\sum_{\alpha} |C_{\alpha}|^4$. This quantity has been shown to increase
in eigenstates of nonintegrable many-body Hamiltonians as one departs from an integrable 
point \cite{santos_rigol_10a}. On the other hand, the fact that the weight of the initial 
state within the microcanonical window decreases if the value of $t',V'$ becomes too large 
can be quantified calculating $W=\sum_{\alpha'}|C_{\alpha'}|^2$, where only eigenstates 
$\alpha'$ inside the microcanonical window are added.

Then, one can compute a ``normalized IPR,'' which is the product of the IPR and $W$. For
finite systems, this quantity can tell us how effective $t',V'$ are in sampling states
within the microcanonical window. Results for this quantity are depicted in 
Fig.~\ref{NormalizedIPRvstpVp}, for the same quenches depicted in Fig.~\ref{nk_DiagvsMic}.
Figure~\ref{NormalizedIPRvstpVp} shows that, in the region where $\Delta n_k$ exhibits a 
sharp decrease in Fig.~\ref{nk_DiagvsMic}, the normalized IPR increases. In 
addition, where $\Delta n_k$ saturates or increases in Fig.~\ref{nk_DiagvsMic},
the normalized IPR saturates or decreases in Fig.~\ref{NormalizedIPRvstpVp}. This allows 
one to understand the overall behavior of $\Delta n_k$ in Fig.~\ref{nk_DiagvsMic}. 
However, we should stress that the fact that 
$t',V'$ (or whatever other term is used to break integrability) cannot be made
too large is only a concern for finite systems. As long as these terms are kept $O(N^0)$
and the interactions have finite range, the energy width of any initial state 
after the quench will vanish in the thermodynamic limit \cite{rigol_dunjko_08}, and the 
initial state will only sample states within the microcanonical window.

{\it Conclusions:} In this Letter we have further probed the role of the eigenstate thermalization 
hypothesis (ETH) as the key dynamical feature of systems that come to thermal equilibrium.  
We have shown that von Neumann's quantum ergodic theorem (QET) relies on a technical assumption 
that is in fact essentially equivalent to ETH.  We have also examined whether, after a sudden quench, 
eigenstates of the initial Hamiltonian can lead to thermal behavior in systems that do not obey 
ETH. We have found that this is possible, but only if those eigenstates obeyed 
ETH before the quench. These results further support the fundamental role of ETH in thermal 
behavior of quantum many-body systems.

\begin{acknowledgments}
MR acknowledges support from the Office of Naval Research. 
MS acknowledges support from NSF Grant No.~PHY07-57035.
We thank M. Kastner for a useful discussion on QET.
\end{acknowledgments}


\begin{thebibliography}{42}
\bibitem{note1} von Neumann actually proved that this ``technical'' condition holds for 
``most'' macroscopic observables. However, ``most'' was defined by a probability 
distribution that treats each observable as a random matrix in the energy basis. 
For real world observables, which are physically motivated and not explicitly built 
as random matrices, that condition has not been proved.
\end{thebibliography}
\end{document}